\documentclass[12pt]{article}
\usepackage{graphicx}

 \usepackage{amsmath}						
 \usepackage{amstext}						
 \usepackage{amsfonts}					
 \usepackage{amssymb}						
 \usepackage{slashed}						
 \usepackage{dsfont}
 
\newcommand\pubnumber{TTP 12-031}
\newcommand\pubdate{August 30, 2012}

\def\Title#1{\begin{center} {\Large #1 } \end{center}}
\def\Author#1{\begin{center}{ \sc #1} \end{center}}
\def\Address#1{\begin{center}{ \it #1} \end{center}}

\newcommand\pubblock{\rightline{\begin{tabular}{l} \pubnumber\\
         \pubdate  \end{tabular}}}
\newenvironment{Abstract}{\begin{quotation}  }{\end{quotation}}
\newenvironment{Presented}{\begin{quotation} \begin{center} 
             PRESENTED AT\end{center}\bigskip 
      \begin{center}\begin{large}}{\end{large}\end{center} \end{quotation}}
\def\Acknowledgements{\bigskip  \bigskip \begin{center} \begin{large}
             \bf ACKNOWLEDGEMENTS \end{large}\end{center}}

\def\beq{\begin{equation}}
\def\eeq#1{\label{#1}\end{equation}}
\def\eeqn{\end{equation}}

\def\beqa{\begin{eqnarray}}
\def\eeqa#1{\label{#1}\end{eqnarray}}
\def\eeqan{\end{eqnarray}}

\let\bar=\overbar

\def\Dslash{\not{\hbox{\kern-4pt $D$}}}
\def\dslash{\not{\hbox{\kern-2pt $\del$}}}

\def\msb{{\bar{\ssstyle M \kern -1pt S}}}

\newcommand{\eins}{\mathds{1}}	
\newcommand{\msbar}{$\overline{\rm MS}$ }
\newcommand{\lvac}{\langle\; {\underset{\raise0.3em\hbox{$\smash{\scriptscriptstyle\thicksim}$}}{0} }\;|}
\newcommand{\rvac}{|\; {\underset{\raise0.3em\hbox{$\smash{\scriptscriptstyle\thicksim}$}}{0} }\; \rangle}

\begin{document}
\begin{titlepage}
\pubblock

\vfill
\Title{Short distance $D^0$ -- $\bar D^0$ mixing}
\vfill
\Author{Markus Bobrowski}
\Address{Institut f\"ur Theoretische Teilchenphysik, Karlsruhe Institute of Technology,\\ D-76131 Karlsruhe, Germany\\markus.bobrowski@kit.edu}
\Author{Alexander Lenz}
\Address{CERN -- Theory Divison, PH-TH, Case C01600,\\ CH-1211 Geneva 23, Switzerland\\e-mail: alenz@cern.ch}
\Author{Thomas Rauh}
\Address{Physik-Department T30d, Technische Universit\"at M\"unchen,\\ D-85748 Garching, Germany\\thomas.rauh@mytum.de}
\vfill
\begin{Abstract}
We review the status of Standard Model predictions for lifetimes and mixing rates of charmed mesons. It is shown that the short distance approach is able to reproduce $\tau({D^+})/\tau({D^0})$ at leading order in the $1/m_c$ expansion. SU(3) violating effects from interactions with the soft hadronic background are identified as the dominant contribution to the $D$--$\bar D$ mixing rate. We discuss the contribution from operators of dimension nine, which is able to enhance the neutral charm width splitting by a factor of order ten and comment on possible CP violation in mixing. 
\end{Abstract}
\vfill
\begin{Presented}
The 5th International Workshop on Charm Physics\\
Honolulu, Hawai'i, May 14--17, 2012
\end{Presented}
\vfill
\end{titlepage}
\def\thefootnote{\fnsymbol{footnote}}
\setcounter{footnote}{0}

\section{Introduction}

Mixing and CP violation of charmed mesons is being probed with unprecedented precision at LHCb \cite{Bediaga:2012py}. Setting bounds of some $10^3\;\rm TeV$ on the effective scale of various $\Delta C = 2$ operators \cite{Isidori:2010kg}, it severely constrains possible extensions of the Standard Model. The $D$ sector is complementary to $B$ and $K$ in offering a handle to probe flavour changing neutral currents among weak isospin up quarks \cite{Gedalia:2009kh}. The recent evidence for CP violation in $D^0 \to K^-K^+,\;\pi^-\pi^+$ decays \cite{DeltaACP} triggered additional interest in charm phenomenology. In the Standard Model, charm physics is dominated by the first two generations, and CP violation is small. In absence of a Standard Model background to interfere with, CP violation in charm is commonly considered as a promising possibility for the search of physics beyond. Large penguin effects, however, partially compromise this feature. 

Flavour oscillations of neutral mesons arise due to non-zero mass and width splittings, $\Delta M$ and $\Delta \Gamma$, between the stationary eigenstates. Charm mixing is experimentally established \cite{DMIX}; the most recent HFAG averages \cite{Amhis:2012bh} for the mixing parameters are 
\begin{equation} 
\label{eq:xyexp}
\begin{aligned} 
  x &\equiv &\frac{{\Delta M}}{\Gamma } &= \left( {0.63^{+0.19}_{-0.20}}
  \right)\% ,\\[.5ex] 
  y &\equiv &\frac{{\Delta \Gamma }} {{2\Gamma }} &=
  \left( {0.75 \pm 0.12} \right)\% .
\end{aligned}
\end{equation}
Two frameworks are being explored to describe these quantities theoretically: the \textit{short distance}, or inclusive, approach assumes quark-hadron duality and is based on parton-level perturbation theory. In the \textit{long distance}, or exclusive, approach the width difference is expressed as a sum over final states common to $D^0$ and $\bar D^0$ decays \cite{EXCL}. In both frameworks, substantial hadronic uncertainties impede definite predictions of the Standard Model expectation. 

Our work is anchored in the inclusive framework. A straightforward application to charm mixing is known to run into severe trouble, what is usually credited to a failure of quark-level perturbation theory at the charm threshold. In the second section below we discuss several issues which we think make it worth to have a closer look, though. We present first results of a Standard Model calculation of the $D^+$--$D^0$ lifetime difference, which avoids the SU(3) flavour interference challenging the calculation of mixing rates.  We find an intriguing agreement with experiment, somewhat clouded by  large parametric uncertainties due to missing lattice input. As regards mixing, SU(3) breaking from hadron state interactions has been conjectured to be the dominant contribution and possibly explain the observed mixing rates within the short-distance picture. We report on a calculation of a specific class of such contributions and find that we can enhance the $D$--$\bar D$ decay width difference by a factor of 
order 10. As already previously, we argue that from the current theoretical status, we can not exclude CP violation in mixing of up to one per cent.

\section{The $D^+$--~$D^0$ lifetime difference}

The heavy-quark expansion \cite{HQE} allows to asses decays of mesons containing one heavy quark. It has proved to be an excellent tool to describe mixing and decays in the $B$ meson sector. The power of heavy-quark techniques became quite impressive earlier this year, when LHCb, CDF, and D0 \cite{EXP} reported on new measurements of the inverse lifetime $1/\tau\left({B_s}\right)$ using angular analysis \cite{Dunietz:2000cr} in $B_s\to \psi\phi$. Contrasting the most recent Standard Model determination of $B$ meson lifetime ratios \cite{LENZNIERSTE} with the numbers presented from LHCb \cite{EXP} in Moriond reveals an amazing agreement: 
\begin{eqnarray}
\frac{\tau\left({B_s}\right)}{ \tau\left({B_d}\right)}_{\rm exp} = 1.001 \pm 0.014\; , &&
\frac{\tau\left({B_s}\right)}{ \tau\left({B_d}\right)}_{\rm SM}  = 0.996 \ldots 1.000 \; .
\end{eqnarray}
In the same analysis, LHCb also reported on the first measurement of a non-zero width splitting $\Delta\Gamma$ in the $B_s$ meson system exceeding $5\sigma$ \cite{EXP}: 
\begin{eqnarray}
  \Delta \Gamma \left(B_s\right) & = &
        \left( 0.116 \pm 0.019 \right) \mbox{ps}^{-1} \; .
\end{eqnarray}
The current HFAG average \cite{Amhis:2012bh} is (this average does not yet include the above number, but uses the one previously published \cite{LHCb:2011aa}): 
\begin{eqnarray}
           \Delta \Gamma\left(B_s\right) & = &
           \left( 0.100 \pm 0.013 \right) \mbox{ps}^{-1} \; .
\end{eqnarray}
These numbers had been of particular interest, since the $B_s$ width difference is believed to be most sensitive to violations of quark hadron duality and receives substantial contributions from hadronic scale dynamics (lattice bag parameters departing from one), perturbative QCD, and subleading $1/m_b$ corrections \cite{Lenz:2011zz}:
\begin{eqnarray}
        \Delta \Gamma\left(B_s\right) & = &  \Delta \Gamma^0\left(B_s\right)\times \left(
         1 + \delta^{\rm lattice} + \delta^{\rm QCD} + \delta^{\rm HQE} \right)
        \nonumber
        \\[1ex]
        & = & 0.142 \; \mbox{ps}^{-1} \left( 1 -0.14 - 0.06 - 0.19 \right) \; .
\end{eqnarray}
Comparing to the most up-to-date Standard Model calculation \cite{LENZNIERSTE},
\begin{equation}
\frac{\Delta \Gamma\left(B_s\right)_{\rm exp}}{\Delta \Gamma\left(B_s\right)_{\rm SM}}
= \frac{0.100 \pm  0.013}{0.087 \pm 0.021} = 1.15 \pm 0.32 \; ,
\end{equation}
shows that, even under more adverse conditions, the heavy-quark expansion managed to provide a solid prediction up to (at least) 30\% accuracy. 

\begin{figure}[t]
\centering
\includegraphics[height=.18\textwidth]{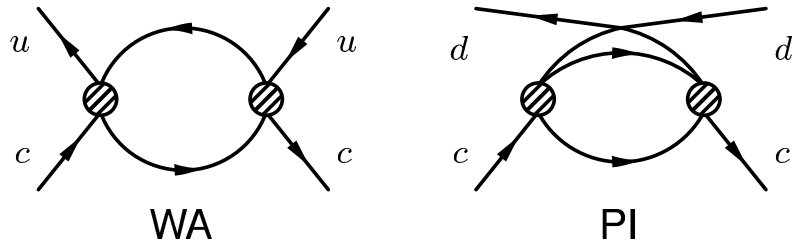}
\caption{\textit{Weak annihilation} (WA) and \textit{Pauli interference} (PI) diagrams at leading order in QCD. They appear in the $1/m_c$ expansion at dimension six and are the dominant contributions to $\Gamma\left(D^0\right)$ and $\Gamma\left(D^+\right)$. Shaded circles indicate insertions of a $\Delta C = 1$ operator.}
\label{fig:diagrams}
\end{figure}

In charm, our image of the inclusive framework is still rather hazy. We do not know whether the radius of convergence in heavy-quark expansion is large enough to allow for a perturbative treatment of mass-suppressed corrections down at the charm threshold. On top of that, the lower scale brings about substantially richer QCD dynamics. Applied to mixing, the naive leading-order predictions in the neutral charm system fall substantially short of the observed values (\ref{eq:xyexp}) for mass and width differences. There are some observations, on the other hand, suggesting that it might be worth to have a closer look at the inclusive short distance picture. To begin with, have a look at what has actually changed compared to the $B$ meson system: heavy-quark expansion is a series expansion in hadronic scale over energy released in decay modes generating the $D$--$\bar D$ or $B$--$\bar B$ transition. The dominant contributions to $\Delta \Gamma$ are from  the $D_s D_s$ final state in $B_s$ and from $KK$ and $\pi\
pi$ in $D$, respectively. It is striking, that the energy releases in $D$ are not significantly smaller than those in $B_s$, where the short-distance toolkit works really well: 
\begin{equation}\nonumber
	\begin{array}{*{20}l}
	   {B_s^0 \; \to \;D_s^ +  \,D_s^ -  } \hspace{2cm} & {1.4\,{\rm{GeV}}}  \\[.6ex]
	   {D^0 \; \to \;\pi \,\pi } & {1.6\,{\rm{GeV}}}  \\[.6ex]
	   {D^0 \; \to \;K\,K} & {0.9\,{\rm{GeV}}}  \\[.6ex]
	 \end{array} 
\end{equation}
With hindsight  we also know that the expansion parameter in $\Delta\Gamma(B_s)$ turned out to be around $1/5$, implying that the relevant hadronic scale is significantly below the $1\,\rm GeV$ it is commonly expected to be. An earlier calculation of subleading corrections in charm mixing \cite{LIFETIMES} likewise did not show signs of a breakdown of the perturbative approach: the charm width difference receives corrections from  next-to-leading order QCD at a level of below 50\%, and $1/m_c$-corrections of 30\%. 

Throughout the last two years, we have advocated \cite{LIFETIMES} that lifetime measurements of charmed hadrons might offer a useful way to assess the applicability of heavy-quark methods to the charm sector on a more quantitative basis: while the leading orders in $1/m_c$ describe spectator model quark decays and are (almost) not sensitive to the meson's light degrees of freedom, $D^0$ mixing and the $D^+$--~$D^0$ lifetime difference arise due to weak interaction with the light valence quarks. They contribute through terms of dimension six onwards and test the heavy-quark expansion at the same order in $1/m_c$. The $D^+$--~$D^0$ lifetime difference, generated by \textit{weak annihilation} and \textit{Pauli interference} (see Fig. \ref{fig:diagrams}), goes back to effects very similar to the ones accounting for the neutral charm width splitting. As opposed to mixing rates and CP violation, yet, it is not an SU(3) breaking quantity and does not suffer from the large GIM suppression inherent to the latter. If 
the heavy-quark expansion is to be applicable, it will have to reproduce charm hadron lifetimes within the first few orders in $1/m_c$, with possible new physics contributions expected to be small. For this, we think that meson lifetimes might be a good test for the inclusive framework in charm.

\begin{table}[t]
\begin{center}
\begin{tabular}{cclr}  
\hline
parameter & input & ~ & rel. error \\ 
\hline
$f_D$ & $206.7\pm8.9\,\rm MeV$ \cite{Rosner:2012bb} & ~ & 5.48\% \\[.5ex]
$B_1$ & $1\pm 1/N_{\rm c}$ & ~ & 20.63\% \\[.5ex]
$B_2$ & $1\pm 1/N_{\rm c}$ & ~ & 0.60\%\\[.5ex]
$\epsilon_1$ & $0 \pm 1/10$ & ~ & 49.47\% \\[.5ex]
$\epsilon_2$ & $0 \pm 1/10$ & ~ & 8.81\% \\[2.5ex]
$\mu_1$ & $1\,{\rm GeV} \ldots 2m_c$ & ~ & $~_{-14.25}^{+\;\;2.50}$\% \\[.5ex]
$\mu_0$ & $1\,{\rm GeV} \ldots 2m_c$ & ~ & $~_{-19.58}^{+11.50}$\%\\[2.5ex]
$\Lambda_{\rm QCD}$ & $222\pm 27\,\rm MeV$   &  ~ & 6.17\%\\[.5ex]
$M_D$ & $1869.60 \pm 0.16 \,\rm MeV$ & ~ & 0.005\% \\[.5ex]
$\bar m_c (\bar m_c)$ & $1.286\pm0.053\,\rm GeV$ & ~ & 4.10\%\\[.5ex]
$\bar m_s (\bar m_c)$ & $0.122^{+0.030}_{-0.039}\,\rm GeV $ & ~ & 0.08\% \\[.5ex]
$m_b$ & $4.651\pm0.054\,\rm GeV$ & ~ & 0.03\% \\[.5ex]
$V_{us}$ & $0.2254\pm0.0013$ & ~ & 0.03\% \\
\hline
\end{tabular}
\caption{Hadronic (top), scale (mid) and experimental (bottom) uncertainties affecting the Standard Model calculation $D^+$--~$D^0$ lifetime difference in the \msbar scheme.}
\label{tab:errorbudget}
\end{center}
\end{table}

\begin{figure}[t]
\centering
\includegraphics[width=.48\textwidth]{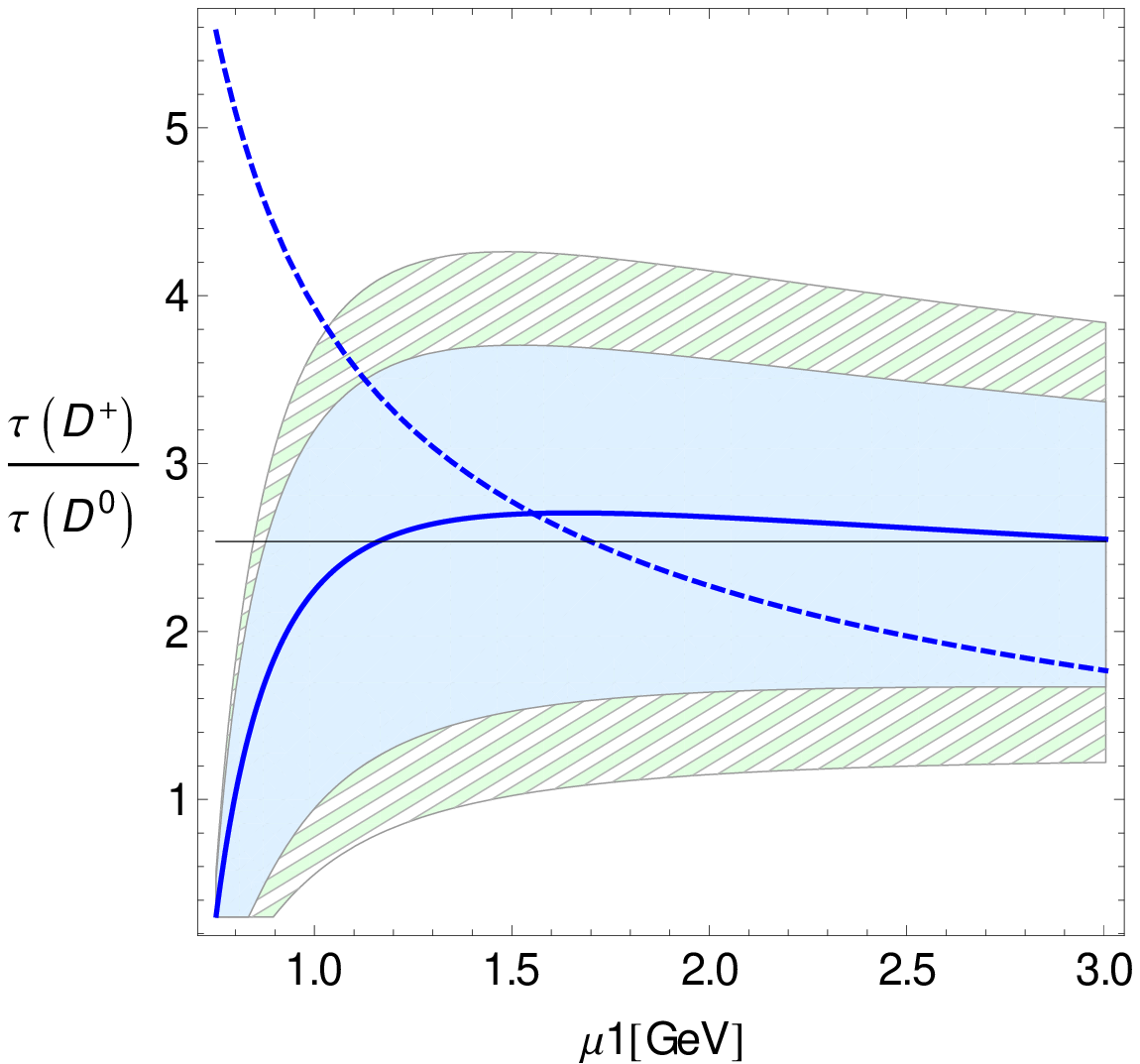}\hspace{.03\textwidth}
\includegraphics[width=.48\textwidth]{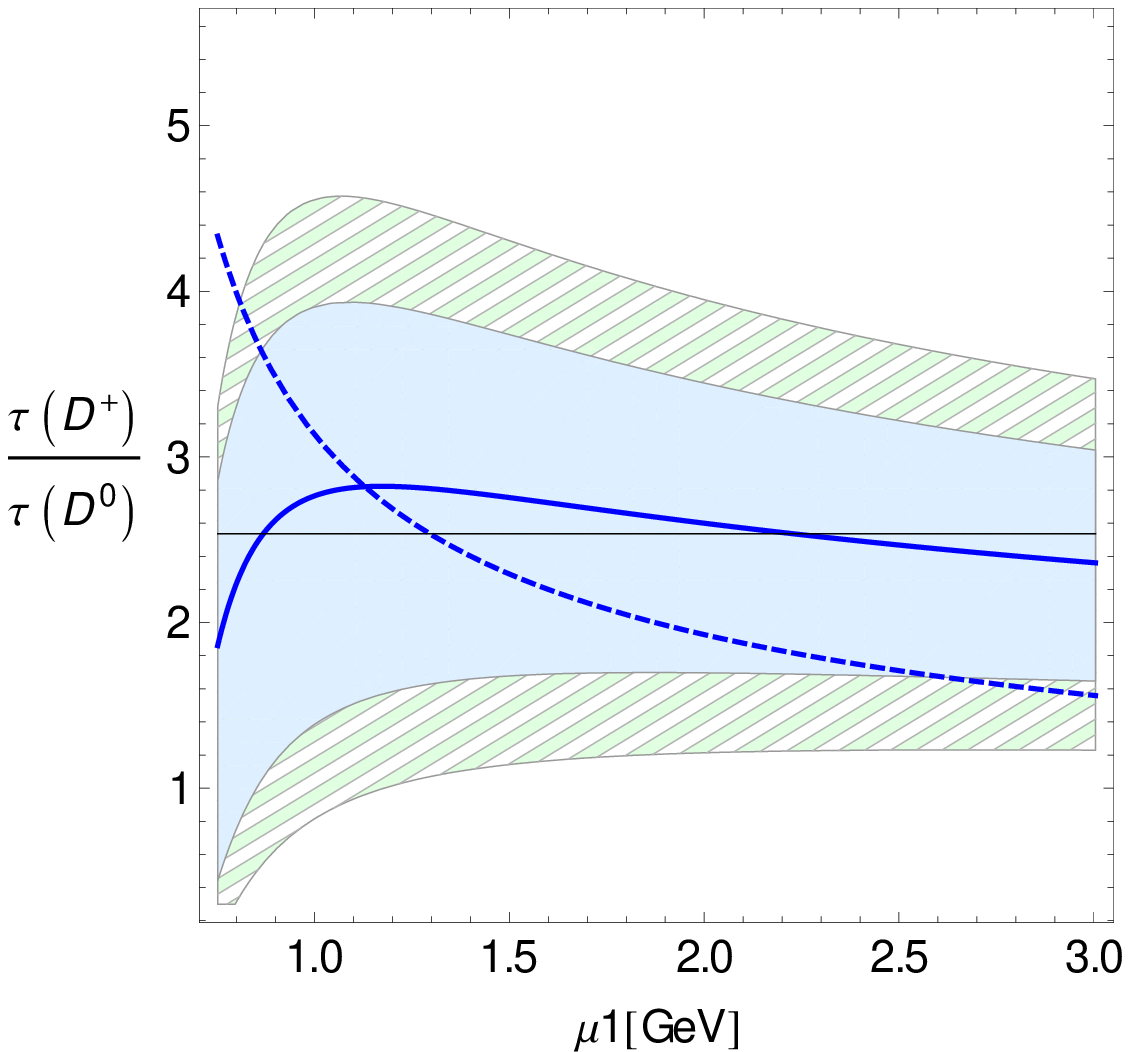}
\caption{The Standard Model estimate for $\tau\left({D^+}\right)/\tau\left({D^0}\right)$ at operator dimension six, as a function of the $\Delta C=1$ renormalization scale $\mu_1$, in the pole (left) and \msbar scheme (right). The plots show the full result at NLO (solid) and the LO only (dashed). The error bands include all sources of uncertainty listed in Tab. \ref{tab:errorbudget}; we display them for a variation of the hadronic parameters $\epsilon_{1,2}$ within $\pm 0.1$ (hatched) and $\pm 0.05$ (shaded). The thin horizontal line marks the ratio of the PDG lifetime averages.}
\label{fig:lifetimes}
\end{figure}

Taking the lifetime ratio $\tau\left({D^+}\right)/\tau\left({D^0}\right)$ at leading order in the $1/m_c$ expansion, and comparing to the naive ratio of experimental averages \cite{Beringer:1900zz}, we find very good agreement: 
 \begin{equation}\label{lifetimes}
 \begin{aligned}
\frac{\tau\left({D^+}\right)}{ \tau\left({D^0}\right)}_{\rm exp} &= 2.536 \pm 0.019\; , \\[.5ex]
\frac{\tau\left({D^+}\right)}{ \tau\left({D^0}\right)}_{\bar{\rm MS}}  &= 2.8 \pm 1.5^{\;\rm (hadronic)} \;_{-0.7}^{+0.3\;\rm (scale)} \pm 0.2^{\;\rm (exp)} \; , \\[.5ex]
\frac{\tau\left({D^+}\right)}{ \tau\left({D^0}\right)}_{\rm pole}  &= 2.7 \pm 1.5^{\;\rm (hadronic)} \;_{-0.9}^{+0.4\;\rm (scale)} \pm 0.2^{\;\rm (exp)} \; .
\end{aligned}
\end{equation}
Our Standard Model estimate is based on results from $B$ physics \cite{Beneke:2002rj} and includes QCD at next-to-leading order, renormalized in both \msbar and pole scheme. Uncertainties due to hadronic and experimental input and the variation of renormalization scale   have been disclosed separately. For the \msbar result, we list their individual sources in Tab. \ref{tab:errorbudget}; to obtain the limits quoted in (\ref{lifetimes}) above, the contributions shown in the Table have been added in quadrature. The dependence on the renormalization scale of the $\Delta C=1$ effective theory is illustrated in Fig. \ref{fig:lifetimes}. Perturbation theory apparently becomes unreliable below about $1\;\rm GeV$, yet it still seems to be under control at the charm threshold. Adding the QCD corrections significantly reduces the scale dependence. To quantify the scale uncertainty in (\ref{lifetimes}), we varied the renormalization scale between $1\;\rm GeV$ and $2m_c$. The overall error is largely driven by hadronic 
uncertainties, entering through matrix elements of the dimension-6, $\Delta C =0 $ operators
\begin{equation}
	\begin{aligned}
	  Q^q \; &= \;\left( {\bar c\,q} \right)_{{\text{V}} - {\text{A}}} \left( {\bar q\,c} \right)_{{\text{V}} - {\text{A}}} , \hfill \\[.5ex]
	  Q_{\text{S}}^q \; &= \;\left( {\bar c\,q} \right)_{{\text{S}} - {\text{P}}} \left( {\bar q\,c} \right)_{{\text{S}} + {\text{P}}} , \hfill \\[.5ex]
	  T^q \; &= \;\left( {\bar c\,T^a \,q} \right)_{{\text{V}} - {\text{A}}} \left( {\bar q\,T^a \,c} \right)_{{\text{V}} - {\text{A}}} , \hfill \\[.5ex]
	  T_{\text{S}}^q \; &= \;\left( {\bar c\,T^a \,q} \right)_{{\text{S}} - {\text{P}}} \left( {\bar q\,T^a \,c} \right)_{{\text{S}} + {\text{P}}} , \hfill \\[.5ex] 
	\end{aligned} 
\end{equation}
where $T^a$ is the generator of colour SU(3). The meson state matrix elements of these operators enter $\tau\left({D^+}\right)/\tau\left({D^0}\right)$ in isospin-breaking combinations, conventionally parametrized as \cite{Beneke:2002rj,blifetime}
\begin{equation}
	\begin{aligned}
	  \left\langle {D^ +  } \right|\,Q^u  - Q^d \,\left| {D^ +  } \right\rangle \; &= \;f_D^2 \,M_D^2 \,B_1 , \hfill \\[.5ex]
	  \left\langle {D^ +  } \right|\,Q_{\text{S}}^u  - Q_{\text{S}}^d \,\left| {D^ +  } \right\rangle \; &= \;f_D^2 \,M_D^2 \,B_2 , \hfill \\[.5ex]
	  \left\langle {D^ +  } \right|\,T^u  - T^d \,\left| {D^ +  } \right\rangle \; &= \;f_D^2 \,M_D^2 \,\varepsilon _1 , \hfill \\[.5ex]
	  \left\langle {D^ +  } \right|\,T_{\text{S}}^u  - T_{\text{S}}^d \,\left| {D^ +  } \right\rangle \; &= \;f_D^2 \,M_D^2 \,\varepsilon _1 . \hfill \\[.5ex] 
	\end{aligned} 
\end{equation}
Due to isospin symmetry, $\langle {D^0 } |\,Q^{u,\,d} \,| {D^0 } \rangle  =\langle {D^ +  } |\,Q^{d,\,u} \,| {D^ +  }\rangle $. In vacuum saturation approximation (VSA), $B_1 = 1$, $B_2 = 1+ \mathcal{O}(\alpha_s,\, \Lambda_{\rm QCD}/m_c)$, while $\epsilon_1$ and $\epsilon_2$ vanish. Non-factorizable corrections to the VSA values are of order $1/N_{\rm c}$.  At the $b$ quark threshold, the hadronic parameters $B_1,\,B_2,\,\epsilon_1,\,\epsilon_2$ are known from lattice QCD \cite{Becirevic:2001fy}. Older studies rely on lattice HQET \cite{DiPierro:1998ty} and HQET sum rules \cite{Baek:1998vk}, and disagree materially with the former. All available non-perturbative studies, though, in agreement with the results for $\tau\left({B^+}\right)/\tau\left({B^0_d}\right)$ \cite{Beneke:2002rj}, consistently indicate that $|\epsilon_1|$ and $|\epsilon_2|$ are significantly smaller than $1/N_{\rm c}$. The results from lattice QCD are, with the information made public, not easily convertible to charm. In our calculation we 
used the VSA values for all hadronic parameters and assigned errors 
\begin{equation}
	\left( {B_1 ,\,B_2 ,\,\varepsilon _1 ,\,\varepsilon _2 } \right)\; = \;\left( {1 \pm {\raise0.7ex\hbox{$1$} \!\mathord{\left/
	 {\vphantom {1 {N_{\text{c}} }}}\right.\kern-\nulldelimiterspace}
	\!\lower0.7ex\hbox{${N_{\text{c}} }$}},\;1 \pm {\raise0.7ex\hbox{$1$} \!\mathord{\left/
	 {\vphantom {1 {N_{\text{c}} }}}\right.\kern-\nulldelimiterspace}
	\!\lower0.7ex\hbox{${N_{\text{c}} }$}},\;0 \pm {\raise0.7ex\hbox{$1$} \!\mathord{\left/
	 {\vphantom {1 {10 }}}\right.\kern-\nulldelimiterspace}
	\!\lower0.7ex\hbox{10}},\;0 \pm {\raise0.7ex\hbox{$1$} \!\mathord{\left/
	 {\vphantom {1 {10 }}}\right.\kern-\nulldelimiterspace}
	\!\lower0.7ex\hbox{10}}} \right).
\end{equation}
A lattice update for these quantities is urgently needed and will shrink the error on $\tau\left({D^+}\right)/\tau\left({D^0}\right)$ drastically. 

Our result (\ref{lifetimes}) certainly supports some confidence in the OPE approach to charm physics. Both the closeness to the observed number and the reasonable behaviour of the QCD perturbation series do not show indications for a breakdown of heavy quark methods. Note, however, that this good agreement still might be by accident; barring significant uncertainties, we so far included only the leading order in $1/m_c$, contributing through operators of dimension six. We expect sizeable corrections from the subleading dimension seven. An estimate for these effects will be crucial to  judge the convergence of the heavy-mass expansion.

\section{Charm mixing and SU(3) breaking}

\begin{figure}[t]
\centering
\includegraphics[height=.18\textwidth]{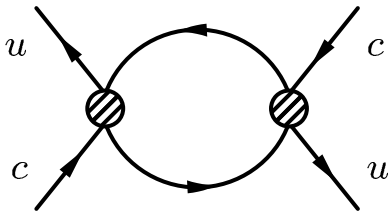}
\caption{The $1/m_c$-leading effect generating a $D$--$\bar D$ transition, contributing to the heavy-quark expansion through operators of dimension six.}
\label{fig:mixing}
\end{figure}

\begin{figure}[t]
\centering
\includegraphics[width=\textwidth]{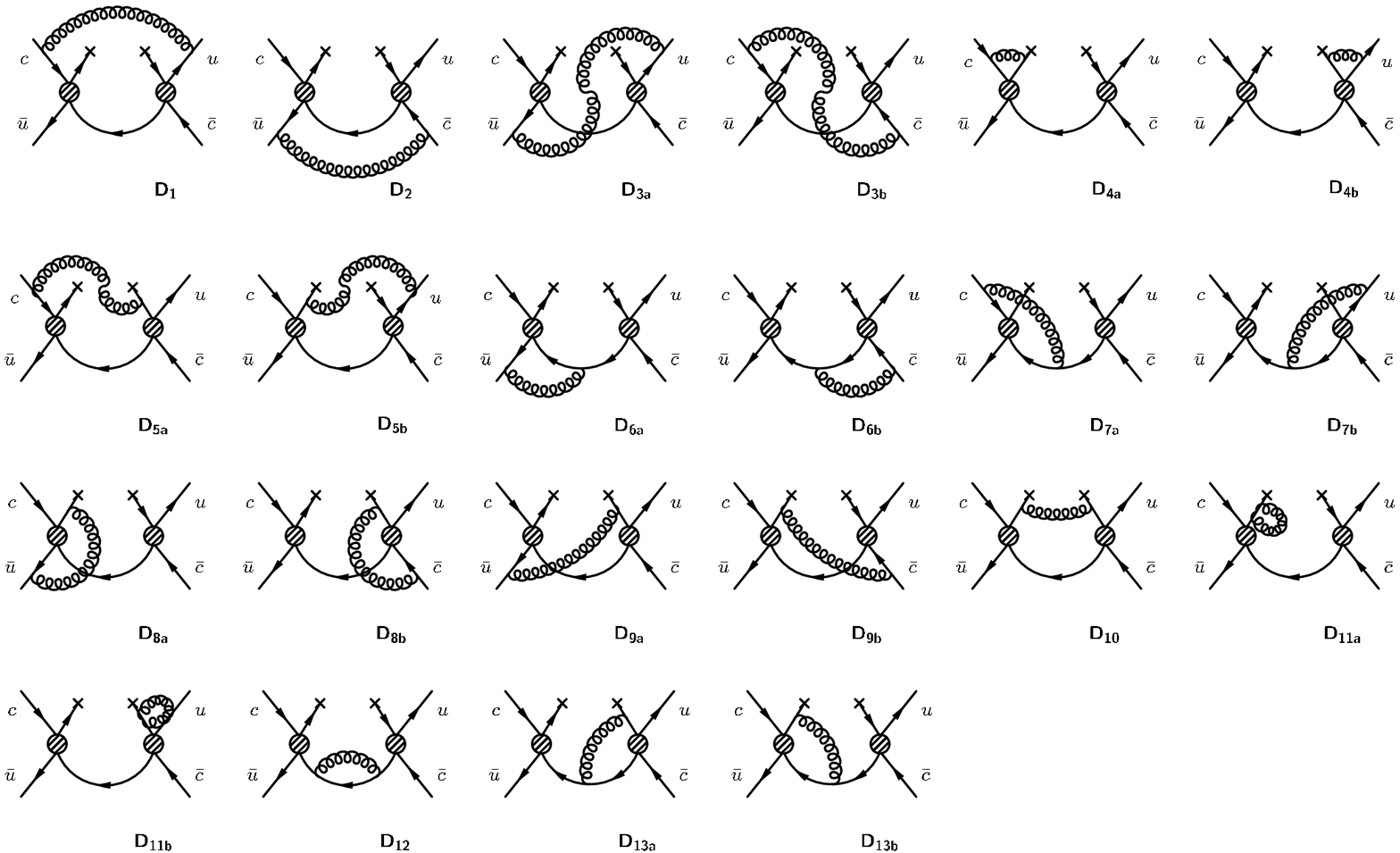}
\caption{$\Delta C =2$ transitions coupling to the hadronic sea quark background in the intermediate state. Shown are the contributions to the $D$--$\bar D$ width difference. One gluon is necessary to generate an on-shell intermediate state.}
\label{fig:cdsLandscape}
\end{figure}

Straightforward application of the heavy-quark expansion to the $\Delta C = 2$ Hamiltonian fails to predict mass and width splittings by orders of magnitude. At leading order in $1/m_c$, charm mixing is predicted to be very slow due to severe GIM suppression. Interference among states in the same SU(3) multiplet almost cancels the  $D$--$\bar D$ transition amplitude. The $D$--$\bar D$ width difference, for example,  receives contributions from on-shell internal $ss$-, $sd$- and $dd$-pairs, see Fig. \ref{fig:mixing}. In the limiting case of exact SU(3), they cancel almost exactly (up to terms of order $|V_{cb}V_{ub}|^2$) as soon as the unitarity of the CKM matrix comes to work. Breaking of SU(3) enters the amplitude through a non-zero strange mass. To overcome the cancellation, one mass insertion per internal line is needed to break SU(3), and a second one to compensate the chirality flip. In the end, a factor of $m_s^4/m_c^4$ (in the CKM-leading part), reminiscent of broken SU(3) symmetry, is suppressing the 
mixing rate. The interference will be prevented as soon as SU(3) breaking is introduced from sources other than the quark masses.  If one of the internal momenta is less than $\Lambda_{\rm QCD}$, the intermediate state couples to the meson's soft QCD substructure and feels the breaking of SU(3) in the hadron state. In the OPE picture, the effect is generated by operators of dimension nine (Fig. \ref{fig:cdsLandscape}). Albeit suppressed by three additional powers of $1/m_c$ with respect to the leading dimension six, it is of lower order in $m_s$ and has been suspected to dominate the heavy-quark expansion \cite{SU3}. Breaking of SU(3) symmetry in non-perturbative hadronic QCD dynamics can access the $\Delta C =2$ Hamiltonian through matrix elements of 6-quark operators containing a $s$- or $d$-quark pair. To estimate the  hadronic matrix elements, we assume that the fields from the intermediate state factorize from the operator structure. With the coupling to the sea quark background dominated by low 
energies, we model the hadronic $s$ and $d$ quark content with the vacuum condensate and evaluate the matrix elements of the factorized 2-quark structure assuming vacuum saturation \cite{Shifman:1978bxyw}. Corrections to factorization are of order $1/N_{\rm c}$. 
The condensate of quark-antiquark pairs in the QCD vacuum is
\begin{equation}\label{eq:defCds}
	\text{
	\raisebox{-8pt}{\parbox[c]{30mm}{
	     \includegraphics[width=30mm]{./eqn.eps}
	}
	}}
\,=\,\lvac \,q\left( x \right) \otimes \overline  q\left( 0 \right)\,\rvac =  - \frac{{\left\langle {\overline  qq} \right\rangle }}
{{4N_{\rm c}}}\, \times \,\eins_{\text{c}} \;\left( {\eins  - \frac{{{\text{i}}\,m_q}}
{4}\;\slashed{x}} +\ldots\right),
\end{equation} 
for our purpose with $q =d,\,s$. The second term on the right-hand side corresponds to the first order term of a Taylor expansion in a small momentum flow through the hadron state. Higher orders in the expansion are suppressed by even more powers of $m_q$ and can be neglected. For the diagrams of Fig. \ref{fig:cdsLandscape} to overcome chirality suppression, accordingly, we have to supply one mass insertion from the right-hand side of (\ref{eq:defCds}), or from the perturbative quark propagation. Taken together, this amounts to a relief in SU(3) interference by one factor of $m_s/m_c$. If this relief is strong enough to outweigh the suppression by additional powers of $1/m_c$ (which isn't actually a so small number), diagrams of this topology would actually dominate the heavy-quark expansion. 

In our calculation of the contributions to $\Delta \Gamma$ as shown in Fig. \ref{fig:cdsLandscape}, we can confirm the expectation from this power-counting. We found that, as expected, the correction to each individual SU(3) amplitude is small (in the percent range), as it should be if the $1/m_c$ expansion is to converge. Due to less pronounced residual flavour symmetry, however, it survives the cancellations affecting the $1/m_c$-leading contribution when differently-flavoured intermediate states are added, and exceeds the latter by a factor of order ten. Likewise, the Standard Model prediction for the $D$--$\bar D$ width difference is enhanced by a factor of order ten. A similar calculation has been done for the $D$--$\bar D$ mass difference. Our results for $x$ and $y$ are
\begin{equation}
\begin{gathered}
  x = \left( {6 \pm 2} \right) \cdot 10^{ - 5} , \hfill \\[1ex]
  y = \left( {8 \pm 9} \right) \cdot 10^{ - 6} . \hfill \\ 
\end{gathered} 
\end{equation}
The mixing rates still miss the experimental numbers (\ref{eq:xyexp})  by a factor around 100 and 1000, respectively. Nevertheless, the calculation shows that the SU(3) breaking hadronic state interactions in dimension nine indeed are the (so far) dominant effect generating the $D$--$\bar D$ decay width difference. The calculation also allows to extract the weak phase in mixing. Up to operator dimension nine, we find $\phi = 1.8_{-0.1}^{+0.2}$. To estimate the physical amount of CP violation mixing might account for, assume that there is some mechanism, able to break also the yet remaining soft SU(3) interference, which originates from the continuing quark lines in Fig. \ref{fig:cdsLandscape}, and from the mass-dependence of the QCD vacuum condensate (\ref{eq:defCds}). The effect will mainly work in amplitudes with real CKM-couplings and enhance the mixing rates while reducing the weak phase.  If we further assume that it is powerful enough to push $x$ and $y$ to their average experimental values, we find 
that a weak phase in the range of one per mille up to (at most) one per cent might be accommodated.

\section{Summary and outlook}

Among the four neutral meson systems $K$, $B$, $B_s$ and $D$, charm is the most challenging to approach theoretically. In these proceedings, we gave a review on Standard Model calculations of the $D$--$\bar D$ mixing rate. Weak interactions of mesons can be described in the framework of the heavy-quark expansion, which had great success predicting lifetimes, mixing and CP violation of $B^0$ and $B^\pm$ mesons with high accuracy. At Moriond 2012, LHCb presented the first measurement ($>5\sigma$) of the decay width difference $\Delta\Gamma$ of neutral $B_s$ mesons, which had been expected to be most sensitive to violations of duality. The agreement with the Standard Model prediction is intriguing, a fortiori in that the energy released in $B_s \to D_sD_s$, which dominantly generates the width splitting, is only around $1.4\,\rm GeV$. Applied to charm, it is well known that the heavy-quark approach meets serious problems. Predicted $D$ meson mixing rates typically fall short of the experimental averages by 
several orders of magnitude, which is commonly attributed to a failure of quark-level perturbation theory at the charm threshold. Some observations, though, argue that it might be possible to understand also the $D$--$\bar D$ width splitting in the Standard Model. Energy releases in  $D^0$ decays contributing to $\Delta \Gamma$ do not significantly deviate from the $1.4\,\rm GeV$ in $B_s$. A closer inspection of the heavy-quark expansion showed that both NLO and $1/m_c$ corrections are reasonably small to allow for a convergence of the series.

We presented first results on the $D^+$--~$D^0$ lifetime difference in the Standard Model, which are in good agreement with experiment. Corrections from QCD at next-to-leading-order are moderate and significantly reduce the scale dependence. Our result is affected by large hadronic uncertainties associated with the matrix elements of four $\Delta C = 0 $ operators. Future lattice calculations could reduce these uncertainties drastically. We think that our result supports some confidence in the short-distance approach to charm. It will be essential, however, to quantify the effect of $1/m_c$ corrections which we expect to be sizeable. 

Neutral meson mixing is an SU(3) breaking observable; in the Standard Model, the quark masses are the spurions of SU(3) symmetry breaking and mass and width differences scale with powers of $m_s/m_c$. Mixing therefore  is predicted to be very slow at leading order in the heavy-quark expansion. It had been expected that SU(3) breaking interactions with the soft hadronic background, contributing in higher orders of the $1/m_c$-expansion, might be the dominant contributions to neutral charm mass and width splittings and explain the measured $D$--$\bar D$ decay width difference without spoiling the overall convergence of the expansion. We investigated the contribution to $\Delta \Gamma$ from 6-quark operators generating the meson-antimeson transition, which is the $1/m_c$-leading contribution where the effect is at work. We find that it is able to enhance the width difference by a factor of oder 10. On the basis of our results, we argue that we can not exclude CP violation in mixing of up to one per cent. 
Future studies should address the contributions from 8-quark operators, where breaking of SU(3) might be even more significant.

\Acknowledgements
M.B. would like to thank the organizers of \textit{Charm 2012} for the invitation and a pleasant workshop. M.B. is grateful to acknowledge the support by grants of the German National Academic Foundation (Studienstiftung des deutschen Volkes), the State of Bavaria (Bayerische Begabtenf\"orderung), the Elite Network of Bavaria (ENB), and the German Academic Exchange Service (DAAD). This work has been supported by the DFG Research Unit SFB/TR9. A.L. thanks the DFG for the support via the HEISENBERG programme.

\end{document}